\documentclass[twocolumn,showpacs,twoside,10pt,superscriptaddress,prl]{revtex4}

\usepackage{graphicx}
\usepackage{epsfig}
\usepackage{epsf}
\usepackage{amssymb}
\usepackage{amsmath}
\usepackage{amsthm}
\usepackage[colorlinks=true,dvipdfm]{hyperref}

\begin{document}

\title{A Non-Oracle Quantum Search Algorithm and Its Experimental Implementation}
\author{Nanyang Xu,$^{1}$ Jin Zhu,$^{1}$ Xinhua Peng,$^{1}$ Xianyi Zhou,$^{1}$ Jiangfeng Du$^{1\ast}$\\
\normalsize{$^{1}$Hefei National Laboratory for Physical Sciences at Microscale and Department of Modern
Physics,University of Science and Technology of China, Hefei, Anhui 230026, People's Republic of China}\\
\normalsize{$^\ast$To whom correspondence should be addressed; E-mail:  djf@ustc.edu.cn}}

\begin{abstract}
Grover's algorithm has achieved great success. But quantum search algorithms still are not complete algorithms
because of Grover's Oracle. We concerned on this problem and present a new quantum search algorithm
 in adiabatic model without Oracle. We analyze the general difficulties in quantum search algorithms and show how to
solve them in the present algorithm. As well this algorithm could deal with both single-solution and
multi-solution searches without modification. We also implement this algorithm on NMR quantum computer. It is the
first experiment which perform a real quantum database search rather than a marked-state search.
\end{abstract}

\maketitle
Quantum computation is a promising way to solve classical hard problems. Although large-scale quantum hardware
has yet been built, quantum computation model in analog to classical circuit is well developed during the last
few years. Based on this model, several quantum algorithms have been designed to perform classical algorithms
with remarkable speedups. The most splendid one among these is Shor's Algorithm\cite{Shor_algorithm}, which can
factorize a big number using a running time only polynomial in the size of the number, while all known classical
algorithms need a exponential time\cite{QCQI}. Another important algorithm\cite{Grover_search}, named after its
inventor Grover, concerns the problem of searching for a required item in a unsorted database. One common example
for this unsorted database search is to find a person's name in a phone book (the items are sorted by names) with
only knowing his phone number. Classically, the only way to achieve this is brute-force search\cite{Ju_2007}
which for $N$ entries in the phone book requires an average of $ \frac{N}{2}$ quires. However, if the information
is stored in a quantum database, to find the right name with Grover's algorithm costs only a time of order
$\sqrt{N}$, providing a quadratic speedup.

While quantum algorithms are presented in standard circuit
model(\textit{i.e.}, using a sequence of discrete quantum gates), a
new model of quantum computation show up where the states of quantum
computer evolves continuously and adiabatically under a certain
time-dependent Hamiltonian. This new idea was firstly brought out by
Farhi and co-workers\cite{Farhi}. In this new computation model, a
problem Hamiltonian is well designed whose ground state encodes the
unknown solution to the problem. Then this adiabatic evolution can
be used to switch gradually from an initial Hamiltonian whose ground
state is known, to the problem Hamiltonian. If this evolution
evolves slowly enough, the system will stay near its instantaneous
ground state\cite{Messiah}. So in the end of evolution, the system
will on the solution state of the problem. This method has been
applied to the database search problem. However, this adiabatic
search algorithm results in a complexity of order $N$, which is the
same order with classical algorithms. More recently, Roland and
Cerf\cite{Roland_2002} improved the performance of adiabatic search
to order $\sqrt{N}$, the same with Grover's algorithm, by applying
adiabatic evolution locally.

Although these quantum search algorithms seems brilliant as they
have already done, they are still incomplete algorithms. Grover's
algorithm utilized a Oracle (\emph{i.e.}, a blackbox) , which gets
an input state $\arrowvert i \rangle$, checks the quantum database,
and changes the state to $-\arrowvert i\rangle$ if the $i$-th value
in the database satisfies the search condition and does nothing
otherwise. It is easy to implement such operations in classical
database cases, but up to now there's no efficient universal method
to design this Oracle in quantum circuit. And in adiabatic
algorithms, the solution of the problem is encoded to the problem
Hamiltonian. Since the mechanics of the Oracle remains unknown, the
encoding process of the Hamiltonian in the adiabatic algorithm is
unclear. Instead, just like what previous
experiments\cite{Chuang_1998search, Dodd_2003,
Vandersypen_3bitsearch, Brickman_2005} of Grover's algorithm did,
the adiabatic search algorithm forms the Hamiltonian directly from
the solution state, which means we have to know the state \emph{in
prior} and then perform a algorithm to show it. Obviously this
marked-state search algorithm is not a real database search. Thus
the main problem with current quantum search algorithms is the
existence of Grover's Oracle.

In this article, we present a new adiabatic algorithm for quantum
search. By encoding the database to quantum format and forming the
problem Hamiltonian from target value, this adiabatic search
algorithm solves Grover's problem without Oracles. Furthermore, we
experimentally implement this non-Oracle quantum search algorithm in
NMR quantum computer. Because of the reasons mentioned before, this
is the first time implementing a real quantum unsorted database
search in experiment. We also analyze the general difficulties in
quantum search and show how to solve them in our algorithm.


As a new quantum computation model, adiabatic algorithm was brought out by Fahi \textit{et al.}\cite{Farhi} and
soon became a rapidly expanding field. This new computing model relies on the \textit{Adiabatic Theorem} which
states as follows: \newtheorem*{adthm}{Adiabatic Theorem}
\begin{adthm}
A physical system remains in its instantaneous eigenstate if a given perturbation is acting on it slowly enough
and if there is a gap between the eigenvalue and the rest of the Hamiltonian's spectrum.
\end{adthm}
The idea of adiabatic quantum computation is straightforward. First
find a complex Hamiltonian whose ground state describes the solution
to the problem of interest, Next, prepare a system with a simple
Hamiltonian and initialized to the ground state. Finally, the simple
Hamiltonian adiabatically switches to the complex Hamiltonian.
According to adiabatic theorem, the system stays in the ground
state, so in the end the state of the system describes the solution
to the problem. The time dependent Hamiltonian is usually
constructed as follows,
\begin{equation} \label{ht}
 H(t)=[1-s(t)]H_{i}+s(t)H_{p},
\end{equation}
where $H_{i}$ is the initial Hamiltonian whose ground state is easy
to know and $H_{p}$ is the complex problem Hamiltonian whose ground
state describe the solution to our problem, and the monotonic
function $s(t)$ fulfills $s(0) = 0$ and $s(T)=1$.

Several adiabatic algorithms have been designed for solving
computational hard problems\cite{Farhi_2000,Peng_2008}. And a simple
proof has been given to show that adiabatic model is equivalent to
circuit mode in quantum computation\cite{Mizel_2007}. Moreover,
since adiabatic computation only involves the ground state, it keeps
the system at a low temperature. Thus the system appears lower
sensitive to some perturbations and have a improved robustness
against dephasing, environmental noise and some unitary control
errors\cite{Childs_2002,Roland_robust}.

As mentioned before, the key part of an adiabatic algorithm is how
to describe the solution to a specific problem in the problem
Hamiltonian. Here let's focus on the database search problem. To be
simplified, we now consider a phone book which contains $N$ (assume
$N = 2^{n}$) entries with each entry a pair of telephone number and
person's name. Usually, the entries are sorted by name. The database
search problem here is to find a specific name in the book whose
telephone number is given. To solve this problem in our model, the
names are encoded to  $n$-qubit states and the phone numbers
represented as integers (in fact, any real numbers are permit). An
example for $N=4$ is shown in Table \ref{phone-book}. We encode the
names and the phone numbers as in Table \ref{encoder}. Thus the
database could be stored by the state-integer pairs like
$\{(|0\rangle,4),(|1\rangle,3),(|2\rangle,1),(|3\rangle,2)\}$. If we
want to find the name which connect to the number 3601003 which is
encoded as 3, state $|1\rangle$ should be returned from our quantum
search machine.

\begin{table}[h]
\caption{phone book example} \label{phone-book}
\begin{center}

\begin{tabular}{ccc}

\multicolumn{1}{c}{} &Name& Number \\
\hline

&Alex & 3601004 \\
&Bob & 3601003 \\
&Cherry & 3601001\\
&David & 3601002 \\

\hline
\end{tabular}
\end{center}
\end{table}

\begin{table}[h]
\caption{encodor table} \label{encoder}
\begin{center}

\begin{tabular}{ccc||cc}

\multicolumn{1}{c}{} & Name &  State  &  Number  &  Integer  \\
\hline

&Alex & $|0\rangle$ & 3601001 & 1 \\
&Bob & $|1\rangle$ &3601002 & 2 \\
&Cherry & $|2\rangle$ &3601003 & 3 \\
&David & $|3\rangle$ &3601004 & 4 \\

\hline
\end{tabular}
\end{center}
\end{table}

After encoding classical database to quantum database, next step is to design the problem Hamiltonian $H_{p}$.
For this problem, $H_{p}$ is.
\begin{equation}
H_{p} = (\sum_{i=0}^{N-1} number_{i}|name_{i}\rangle \langle name_{i}|-target \cdot I ^{\otimes n})^{2}
\end{equation}
where \textit{target} is the code of the phone number which we want to search for. Obviously, the ground state of
$H_{p}$ is the state which connects to the target number. In general, for an encoded database where entries are
pairs as $(i,value_{i})$ and sorted by $i$, we write $H_{p}$ as
\begin{eqnarray}
\label{hpd}
H_{p} &=& (\sum_{i=0}^{N-1} value_{i}|i\rangle \langle i| - target\cdot I ^{\otimes n})^{2}\nonumber\\
    &=& \mathcal{D}^{2}-2\cdot target\cdot\mathcal{D} + target^2,
\end{eqnarray}
where $\mathcal{D}=\sum_{i=0}^{N-1} value_{i}|i\rangle \langle i|$ is the database operator and could be
formulated separately.

Next, we will choose an initial Hamiltonian $H_{i}$. Conventionlly\cite{Farhi_2000}, $H_{i}$ is chosen to be
noncommutative
 with $H_{p}$ to avoid crossing of energy levels. Thus we write $H_{i}$ as:
\begin{eqnarray}
H_{i} & = & g(\sigma_{x}^{0}+\sigma_{x}^{1}+\cdots + \sigma_{x}^{n-1}).
\end{eqnarray}
which means the qubits coulpling with a magnetic field at the $x$-direction and the coupling strength is $g$. the
gound state of this Hamiltonian is simple and they are,
\begin{eqnarray}
|\psi_{0}\rangle & = &\frac{|0^{(n-1)}\rangle-|1^{(n-1)}\rangle}{\sqrt{2}}
\otimes \frac{|0^{(n-2)}\rangle-|1^{(n-2)}\rangle}{\sqrt{2}} \otimes \cdots \nonumber\\
&\otimes&  \frac{|0^{(0)}\rangle-|1^{(0)}\rangle}{\sqrt{2}} \nonumber \\  & =&
\frac{1}{\sqrt{N}}\sum_{j=0}^{N}(-1)^{b(j)}|j\rangle,
 \label{Int_state}
\end{eqnarray}
where $b(j)$ is the Hamming distance between $j$ and $0$.

In the adiabatic evolution, the system Hamiltonian interpolates from $H_{i}$ to $H_{p}$ (\textit{i.e.,} see Eq
\ref{ht}) and the state of the system evolves according to the Schr\"{o}dinger equation:
\begin{eqnarray}
i\frac{d}{dt}|\psi(t)\rangle=H(t)|\psi(t)\rangle  \\
|\psi(0)\rangle=|\psi_{0}\rangle \label{e.Schord}.
\end{eqnarray}
If this evolution acts slow enough (\textit{i.e.,} the total evolution time $T$ is large enough), the
\textit{\textbf{Adiabatic Theorem}} ensures the system will always stay on the ground state of $H(t)$ and in the
end the solution of our problem will show up.

\begin{figure}[t]
\begin{center}
\includegraphics[width= 1\columnwidth]{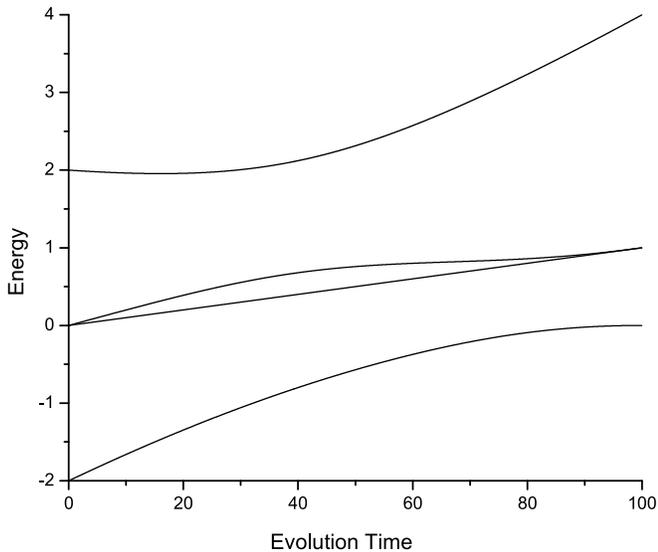}
\end{center}
\caption{Energy diagram for searching an item (\textit{i.e.}, 3601003) in the example database (see Table
\ref{phone-book}). During the adiabatic evolution, the state will remain on the lowest energy level.}
\label{energy}
\end{figure}

Again, we take the phone book in Table \ref{phone-book} as example. If we want to find number 3601002 in the
database and using the encoder in Table \ref{encoder}, we will get the Hamiltonians as follows,
\begin{eqnarray}
H_{i} & = & g(\sigma_{x}^{0}+\sigma_{x}^{1})\nonumber\\
H_{p}&=&(4|0\rangle \langle0| + 3|1\rangle \langle1| + 1|2\rangle \langle2| + 2|3\rangle \langle3| - 2I)^{2}  \nonumber\\
&=&\dfrac{3}{2}I +\sigma_{z}^{(0)}+\sigma_{z}^{(1)}+\dfrac{\sigma_{z}^{(1)}\otimes \sigma_{z}^{(0)}}{2},
\label{david}
\end{eqnarray}
And the eigenvalues of time dependent Hamiltonian $H(t)$ (see EQ.(\ref{ht}) ) are plotted in FIG.\ref{energy}. By
the adiabatic theorem, the state will stay on the lowest energy level during the adiabatic evolution. And finally
we will get the state on the basis $|3\rangle$. After measurement and decoding we will get the name connecting to
the number 3601002 which is \textit{David}.

To demonstrate this Non-Oracle search algorithm, we selected $^{13}$C-labeled CHCl$_3$ as a physical system for
our experiments. The two qubits are represented by $^{13}$C and $^1$H. Its natural Hamiltonian in the multiply
rotating frame is
\begin{equation}
\mathcal{H}_{sys}=\omega_1 I_z^1+\omega_2 I_z^2+2\pi J I_z^1 I_z^2,
\end{equation}
where $\omega_1$ and $\omega_2$ are Larmour frequencies, J is the spin-spin coupling constant $J=214.5Hz$.
Experiments were performed at room temperature using a standard 400MHz NMR spectrometer (AV-400 Bruker
instrument).

\begin{figure}[tbph]
\centering
\includegraphics[width=0.8\columnwidth]{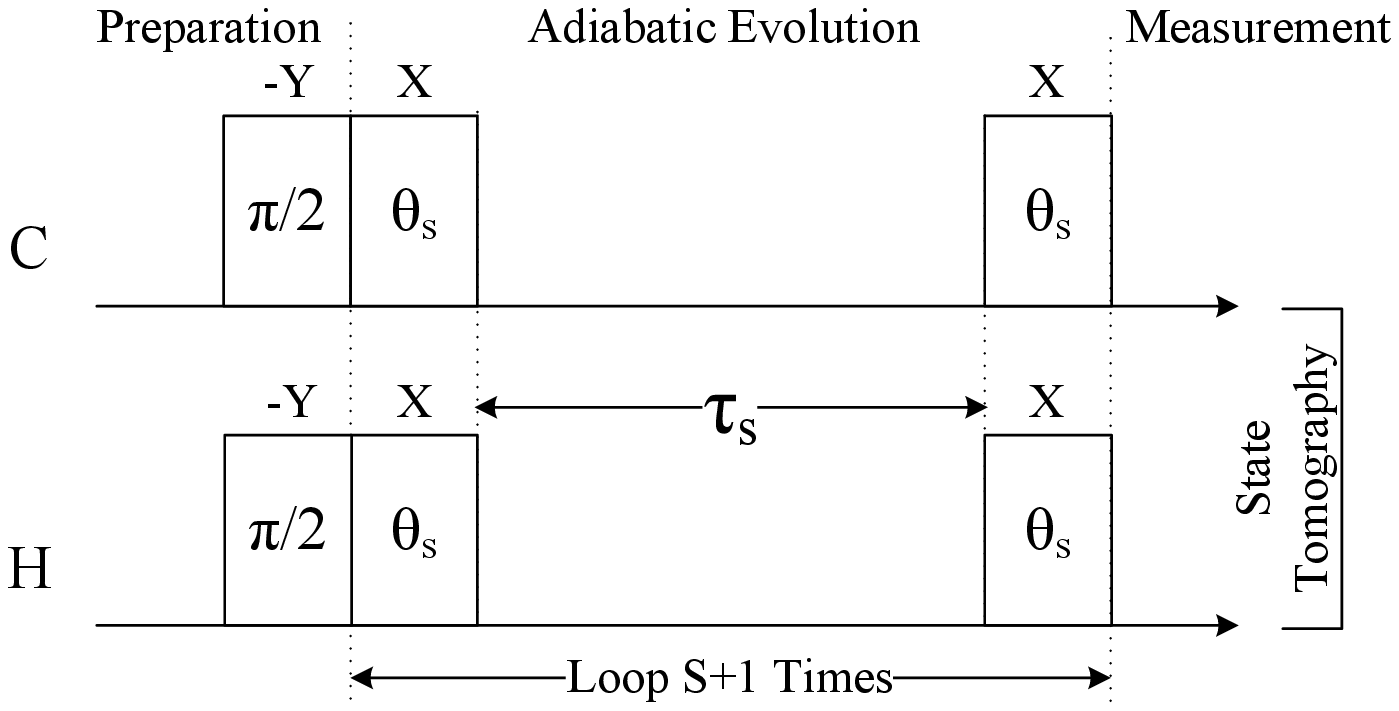}
\caption{The quantum network for adiabatic search using NMR interferometry. The input state is $|00\rangle$.
$\theta_s=0.95*(1-s/10)$, $\tau_s=0.95*(s/10)/\pi J$. } \label{net}
\end{figure}

The experiments was devided into three parts, shown as Fig.\ref{net}: the first part consists of preparation of
the state of the initial Hamiltonian. The second part is the adiabatic evolution, and the third part is the
tomography of the resultant state. To prepare the initial ground state [Eq. \ref{Int_state}], we first created a
pseudopure state (PPS) \cite{Gershenfeld:1997aa,Cory:1997aa} $\rho _{00}=\frac{1-\epsilon }{4}\mathbf{I}+\epsilon
|00\rangle \langle 00|$, where $\epsilon \approx 10^{-5}$ describes the thermal polarization of the system and I
is a unit matrix, using the method of spatial averaging. Then $\pi/2$ pulses along the $-y$ axis was applied to
prepare the ground state.

Discretizing a continuous Hamiltonian is a straightforward process and changes the run time T of the adiabatic
algorithm only polynomially. Simply, let the discrete time Hamiltonian $H(s)$ be a linear interpolation from some
beginning Hamiltonian $H(0)=H_0$ to some final problem Hamiltonian $H(S)=H_1$ such that
$H(s)=(s/S)H_1+(1-s/S)H_0$. The unitary evolution of the discrete algorithm can be written as
\begin{equation}
U=\prod_{s} U_s=\prod_{s} e^{-iH(s)\tau},
\end{equation}
where $\tau=T/(S+1)$, $T$ is the total duration of the adiabatic passage and $S+1$ is the total number of
discretization steps. When both $T, S \to \infty$ and $\tau \to 0$, the adiabatic limit is achieved.

For our example shown in Eq.\ref{david}, an optimized set of parameters was set as $g=1$, $T=10.45$ and $S=10$,
so $\tau=T/(S+1)=0.95$. This set of parameters yields an adiabatic evolution that finds the solution in a
relatively efficient way. Using the Trotter formula, we can approximate $U_s$ to second order
\begin{equation}
U_{s} \approx U'_s= e^{-i (1-\frac{s}{S})H_0\frac{\tau}{2}} e^{-i\frac{s}{S}H_{p}\tau} e^{-i
(1-\frac{s}{S})H_0\frac{\tau}{2}} + \mathcal{O}(\tau ^2), \label{trotter}
\end{equation}
the fidelity of $U_s \to U'_s$ is all above $0.996$ and overall fidelity is $0.991$. For the implementation of
$U'_s$, $e^{-i (1-\frac{s}{S})H_0\frac{\tau}{2}}$ can be simply realized using a $\theta_s$ pulses around $x$
axis for both H and C nuclei, $\theta_s=\tau*(1-s/S)$, shown in Fig. \ref{net}. And the evolution under $H_1$ can
be simulated by a free evolution $\tau_s$ under the Hamiltonian $\mathcal{H}_{sys}$, the identity term of $H_1$
does not cause any evolution of the state and so it can be omitted, $\tau_s=\tau*(s/S)/\pi J$.

\begin{figure}[tbph]
\centering
\includegraphics[width=\columnwidth]{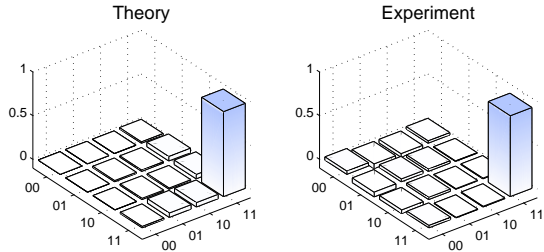}
\caption{The tomography of theoretically expected and experimentally obtained density matrices for the search
states in adiabatic search algorithm. The density matrices consist of just a real term on the diagonal
corresponding to the population of the state that has been searched.} \label{result}
\end{figure}

The third stage of the experiment is the tomography of the final density matrix after the adiabatic evolution.
The result was shown as Fig. \ref{result}. Theoretically, the four state $|00\rangle$, $|01\rangle$, $|10\rangle$
and $|11\rangle$ should be find at the probability of 0, 0.014, 0.014 and 0.972. Our experiments show that the
probability is 0.037, 0.032, 0.006 and 0.925. The fidelity of the experiment is 0.985.

The errors in adiabatic algorithms may be caused by three parts. Firstly, the total time of evolution in
adiabatic algorithms should be infinite. Actually the evolution is terminated when the state is supposed to reach
our expected high probability. Secondly, the error is due to neglect of $\mathcal{O}(\tau^2)$ terms in the
Trotter Formula (Eq. \ref{trotter}). The third part of error is due to decoherence effects of the NMR system and
imperfect pulses.

Unlike previous standard quantum algorithms only using qubits as
registers to store information, our algorithm represents the $value$
field by the strength of interactions in the operator and the
$index$ field by qubits. This is because that if both the fields are
represented in qubits, $2n$ qubits are needed for a database with
$N$ items, which result in the failure that the optimal running time
would be scaled from order $\sqrt{N}$ to $N$\cite{Roland_2002}, the
same performance as classical algorithms. Since the construction
function is simply quadratic, the interaction strength in problem
Hamiltonian grows with the database's size. For further
consideration, the algorithm may be improve to suppress the
interaction strength in the problem Hamiltonian by choosing a better
construction function in Eq. \ref{hpd}.

All practical quantum search algorithms must face the problem that
the database is unsorted, thus quantum operators would traverse all
the items in the database to learn the complete information and
after measurement all information in the states are destroyed. It is
a hard problem to efficiently implement the quantum operators
concerning the database. The first effort was reported by Ju and
coworkers\cite{Ju_2007} when they tried to implement Grover's Oracle
in quantum circuit. They have to spend $N$ steps to build the
relation in the database to the states for each query, such that the
circuit design is not efficient. In this algorithm, we describe the
database information in a single operator($\mathcal{D}$ in
Eq.\ref{hpd}), thus this operator may be analyzed and formulate
separately for each database. On the other side, if it could not be
formulated efficiently in some cases, approximate implementation is
another possible solution according to recent works on geometric
quantum computation\cite{Nielsen_geometry}, of which detailed
consideration is beyond the scope of this article.

As an end of this section, we will give a simple analysis of the multi-solution search case in our algorithm. If
there're $m > 1$ entries in the database satisfying the search condition, the problem Hamiltonian will have
ground states with $m$ times degenerated. And because of symmetry, the state would finally evolute to an average
superposition of all the ground states. Thus without any modification, our algorithm could also deal with
\emph{multi-solution search}.

To be concluded, we present a new kind of adiabatic search algorithm
to solve Grover's problem without Oracles and give a demonstrative
experiment on NMR quantum computer. The result of experiment agrees
well with theoretical expectation. This is a new style of quantum
search algorithm which utilize both quantum registers and
interaction strength to store information. This algorithm aims at
general difficulties of quantum search algorithms and give a
promising way to solve them utimately.

We thank Zeyang Liao for initial discussion and help. This work was
supported by National Nature Science Foundation of China, the CAS,
Ministry of Education of PRC, the National Fundamental Research
Program, and the DFG through Su 192/19-1. For this article, any
comment is welcome.

\bibliography{adiabatic_quantum_computing}

\end{document}